\documentstyle[12pt]{article}
\newcommand{\thspace}{\kern.08333em}

\def \beqn{\begin{eqnarray}}
\def \beq{\begin{equation}}

\def \eeq{\end{equation}}
\def \eeqn{\end{eqnarray}}

\def \s{\sqrt{2}}

\begin{document}

\rightline{SLAC-PUB-8216}
\rightline{hep-ph/9908237}
\rightline{August 1999}

\bigskip
\bigskip
\begin{center}
\Large\bf \boldmath Resonant Two-body $D$ Decays
\unboldmath
\small\footnote{Supported
in part by the Department of Energy under contract number DE-AC03-76SF00515.}
\end{center}
\bigskip
\bigskip
\centerline{Michael Gronau~\footnote{Permanent Address: Physics
Dept.,
Technion -- Israel Institute of Technology, 32000 Haifa, Israel.}}
\centerline{\it Stanford Linear Accelerator Center}
\centerline{\it Stanford University, Stanford, CA 94309}
\bigskip
\bigskip

\centerline{\bf ABSTRACT}
\bigskip

\begin{quote}
The contribution of a $K^*(1430)$ $0^+$ resonance to
$D^0\to K^-\pi^+$ is calculated by applying the soft pion
theorem to $D^+ \to K^* \pi^+$, and is found to be about
30$\%$ of the measured amplitude and
to be larger than the $\Delta I=3/2$ component of this amplitude.
We estimate a 70$\%$ contribution to the total amplitude from a higher
$K^*(1950)$ resonance. This implies large deviations from factorization
in $D$ decay amplitudes, a lifetime difference between $D^0$ and $D^+$,
and an enhancement of $D^0-\bar D^0$ mixing due to SU(3) breaking.
\end{quote}
\bigskip
\bigskip
\bigskip
\bigskip
\begin{center}
To be published in Physical Review Letters
\end{center}

\newpage

Hadronic two-body and quasi two-body weak decays of $D$ mesons, which
constitute a sizable fraction of all hadronic $D$ decays \cite{ST,WI},
involve nonperturbative strong interactions.
Long distance QCD effects spoil the simplicity of the short distance
behavior of weak interactions \cite{GLAM}. Therefore,
a simplified approach in which
the amplitudes of these processes are given by a
factorizable short-distance current-current effective Hamiltonian is not
expected to work too well. Various approaches were employed to
include long distance effects. The most commonly and very frequently
used prescription, motivated by $1/N_c$ arguments \cite {FISY}, is to apply
``generalized factorization'' \cite{DGS, BSW, NS, C}: The
two relevant Wilson coefficients ($c_1, c_2$), multiplying appropriate
four-quark short distance operators, are replaced by
scale-dependent free parameters ($a_1, a_2$). In this prescription, the
magnitudes of isospin amplitudes are calculated from experimentally determined
decay constants and form factors, while strong phases (to be determined from
experiment) are assigned to these amplitudes to account for final state
interactions. In spite of its somewhat {\it ad hoc} and disputable procedure
(evidently
final state phases do not occur only in elastic scattering, but are
largely due to inelastic processes), this phenomenological treatment works
reasonably well in Cabibbo-favored $D$ decays \cite{BSW,C}. Its failure in the
Cabibbo-suppressed $D\to\pi\pi$ and $D\to K\bar K$ processes \cite{KP} is
believed to be associated with inelastic hadronic rescattering.

It was pointed out almost twenty years ago \cite{Lip} that the observed
resonance states in the $K\pi,~K\rho/K^*\pi,~\pi\pi,~\pi\rho$ channels,
with masses close to the $D$ mass, may strongly affect final state
interactions in $D$ decays \cite{CL}. The idea is clear and simple,
however its
implementation involves a multi-channel rescattering S-matrix which cannot
be quantified in a model-independent manner \cite{Kam}. In practice, it
is impossible to calculate the effect of s-channel resonance states in
two-body $D$ decays without knowing the weak couplings of the $D$ meson
to these resonances. If some of these couplings are sufficiently
large, the corresponding resonances may have large or even dominating
contributions in certain decays. In this case the apparent success in
describing two-body and quasi two-body decays in terms of ``generalized
factorized'' amplitudes would be an accident which ought to be further
investigated.

Large resonance contributions in $D^0$ decays could explain the observed
$D^+ - D^0$ lifetime difference. Contrary to the $D^0$, the final states in
Cabibbo-favored $D^+$ decays, made of $\bar d s\bar d u$, are pure $I=3/2$ and
do not receive such contributions. Also, resonant amplitudes involve large
SU(3) breaking in the resonance masses and widths. Consequently
intermediate resonance states are expected to lead to large $D^0 - \bar D^0$
mixing. We return to these questions in our conclusion.

The purpose of this Letter is to present the first model-independent
quantitative study of direct channel resonance contributions to two-body
$D$ decays. We will calculate the contribution of
$\bar K^{*0}(1430)$, a particular excited $K$-meson $0^+$ state
($s\bar d$ in a P-wave), to the Cabibbo-favored $D^0 \to K^-\pi^+$ decay
process. In spite of the fact that this resonance peaks at 436 MeV
below the $D$ mass, we find its contribution to amount to a sizable fraction,
approximately 30$\%$,
of the measured $D\to K^-\pi^+$ amplitude. Another $0^+$ $K\pi$-resonance,
observed around 1900 MeV, is likely to have a larger contribution due its
close proximity to the $D$ meson mass. Assuming that its weak coupling to
$D$ is approximately equal to that of the resonance at 1430 MeV,
we estimate its contribution to be about 70$\%$.

An important step in our analysis is the evaluation of the
weak interaction matrix element between a $D$ meson and the 1430 MeV resonance
state. For this purpose, we apply the soft pion theorem which relates this
amplitude to the measured $I=3/2~D^+\to \bar K^{*0}\pi^+$ amplitude
\cite{scadron}.
It is crucial in our argument that the final state $\bar K^{*0}\pi^+$ is
``exotic'', in which
case the amplitude does not involve a pole term (``surface term'') and
varies smoothly and only slightly in the soft pion limit.

The $1430~0^+~K^*$ resonance contribution to $D^0\to K^-\pi^+$ is given
by a Breit-Wigner form
\beq\label{res}
A(1430, K^-\pi^+) = \frac{h_1g}{m^2(D^0)-m^2+im\Gamma}~,
\eeq
where $h_1\equiv <\bar K^{*0}|H_W|D^0>,~m(D^0)=1864.6\pm 0.5~{\rm MeV},~
m\equiv m(K^{*0})=1429\pm 6~{\rm MeV},~\Gamma\equiv \Gamma(K^{*0})
=287\pm 23~{\rm MeV}$ \cite{PDG}.
The strong $K^{*0}K\pi$ coupling $g$ is obtained from the $K^{*0}$ width
\cite{PDG}
\beq\label{g}
g^2=\frac{8\pi m^2\Gamma f}{p_\pi}~,~~~f\equiv {\rm BR}(\bar K^{*0}\to
K^-\pi^+)=0.62\pm 0.07~,~~~p_\pi=621~{\rm MeV}~.
\eeq

The hadronic weak matrix element $h_1$ is related to the measured $I=3/2$
amplitude $h_2\equiv <\bar K^{*0}\pi^+(q_\pi)|H_W|D^+>$ through the
soft pion theorem \cite{Adler}
\beq\label{spl}
\lim_{q_\pi\to 0}<\bar K^{*0}\pi^+(q_\pi)|H_W|D^+>~=~\frac{-i}{f_\pi}
<\bar K^{*0}|[Q^-_5,H_W]|D^+>~,
\eeq
where $f_\pi=130$~MeV and $Q^-_5$ is the axial charge. Note that the
amplitude on the left-hand-side
involves no pole term since $\bar K^{*0}\pi^+$ is an $I=3/2$
state. (On the other hand, the $I=1/2$ $D\to K^*\pi$
amplitude contains such a pole term from an intermediate $0^-(1460)$
$K\pi$ resonance \cite{PDG}, and consequently does not vary smoothly
in the soft
pion limit). The (V-A)(V-A) structure of the $\Delta I=1$ weak
Hamiltonian implies
\beq
[Q^-_5,H_W]~=~-[Q^-,H_W]~,
\eeq
and the isospin-lowering operator $Q^-$
obeys $Q^-|D^+>=|D^0>,~<\bar K^{*0}|Q^- =0$.
Neglecting the small variation in the $D^+ \to \bar K^{*0}\pi^+$ amplitude as
one moves the pion four momentum from its physical value to zero, one finds
\beq\label{h12}
|h_1|~\approx~f_\pi|h_2|~.
\eeq

The amplitude $h_2$ is obtained from the measured width
$\Gamma(D^+\to K^{*0}\pi^+)$ \cite{PDG, CLEO}
$$
h^2_2 = \frac{8\pi m^2(D^+)\Gamma(D^+\to K^{*0}\pi^+)}{q_\pi}~,~~~m(D^+)=
1869\pm 0.5~{\rm MeV}~,~~~q_\pi=368~{\rm MeV}~,
$$
\beq\label{h2}
\Gamma(D^+\to K^{*0}\pi^+)=\frac{0.023\pm 0.003}{\tau(D^+) f}~,~~~
\tau(D^+)=1.051\pm 0.013~{\rm ps}~.
\eeq
Combining (\ref{res})(\ref{g})(\ref{h12})(\ref{h2}), one finds
\beq\label{kstar}
|A(1430, K^-\pi^+)| = (7.85 \pm 0.65)\times 10^{-7}~{\rm GeV}~.
\eeq
The error contains only experimental errors. The uncertainty due to
taking the soft pion limit $q_\pi\to 0$ in the smoothly varying amplitude is
assumed to be smaller and is neglected.
It would be interesting to study this correction, which could slightly
increase or decrease the amplitude.

In order to compare the calculated $K^*(1430)$ resonance contribution to
the measured $I=1/2$ term in $D^0\to K^-\pi^+$, one expresses all three
$D\to\bar
K \pi$ amplitudes in terms of isospin amplitudes. Using a somewhat different
normalization than elsewhere \cite{WI, R1}, we write
\beqn
A(D^0\to K^-\pi^+) &=& A_{1/2} + A_{3/2}~,\nonumber\\
\s A(D^0\to \bar K^0\pi^0) &=& -A_{1/2} + 2A_{3/2}~,\nonumber\\
A(D^+\to\bar K^0\pi^+) &=& 3A_{3/2}~.
\eeqn
Consequently
\beqn
|A_{1/2}|^2 &=& \frac{2}{3}[|A(D^0\to K^-\pi^+)|^2 +
|A(D^0\to \bar K^0\pi^0)|^2 - \frac{1}{3}|A(D^+\to \bar K^0\pi^+)|^2]~,
\nonumber\\
|A_{3/2}| &=& \frac{1}{9}|A(D^+\to \bar K^0\pi^+)|^2~,\\
\cos\delta_I &=& \frac{|A(D^0\to K^-\pi^+)|^2 - 2|A(D^0\to \bar K^0\pi^0)|^2
+\frac{1}{3}|A(D^+\to \bar K^0\pi^+)|^2 }{6|A_{1/2}A_{3/2}|}~,\nonumber
\eeqn
where $\delta_I$ is the relative phase between isospin amplitudes.
One then finds from the experimental rates \cite{PDG, CLEO} the values
\cite{R1}
\beqn
|A_{1/2}| &=& (24.5 \pm 1.2)\times 10^{-7}~{\rm GeV}~,\nonumber\\
|A_{3/2}| &=& (4.51 \pm 0.22)\times 10^{-7}~{\rm GeV}~,\nonumber\\
\delta_I &=& (90\pm 7)^{\circ}~.
\eeqn
This and (\ref{kstar}) imply
\beq
\frac{|A(1430, K^-\pi^+)|}{|A_{1/2}|} = 0.32 \pm 0.03~.
\eeq
That is, the 1430 MeV $K\pi$ resonance contribution is about 30$\%$ of the
dominant $I=1/2$ amplitude in $D\to K\pi$. Its contribution to
$D^0\to K^-\pi^+$ is larger than the $I=3/2$ component of this amplitude.
Note that $A(D^0\to K^-\pi^+)\approx
A_{1/2}$, since $|A_{3/2}|^2\ll |A_{1/2}|^2$ and $\delta_I\approx
90^{\circ}$

In view of this sizable result, which is rather striking for a resonance
peaking 436 MeV below the $D$ mass, one raises the question of
possibly larger contributions to $D\to K\pi$ from resonances lying closer
to the $D$. One such resonance state, around 1900 MeV (denoted $K^*(1950)$ in
\cite{PDG}), was observed in $K\pi$ scattering \cite{LASS}, with
a mass $m'= 1945\pm 22$~MeV and a width  $\Gamma'= 201\pm 86$~MeV. Somewhat
different
values, $m'= 1820\pm 40$~MeV,~ $\Gamma'=250\pm 100$~MeV, were obtained
in a K-matrix analysis \cite{Anis}. Since this resonance lies right at
the $D$ mass, its contribution
to $D^0\to K^-\pi^+$ is likely to be larger than that of $K^*(1430)$.
In order to calculate this contribution, one must know the matrix element
$<K^*(1950)|H_W|D>$, for instance by relating it to $<K^*(1430)|H_W|D>$.
The higher resonance is most likely a radial n=2 excitation
of the state at 1430 MeV which is an n=1 P-wave $s\bar d$ state. In both
amplitudes the local $H_W$ connects a $c\bar u$ S-wave state to
an $s\bar d$ P-wave state which is more spread out. The radially excited
n=2 state is slightly less localized than the n=1 state. Consequently, one
expects $<K^*(1950)|H_W|D>$ to be slightly smaller than $<K^*(1430)|H_W|D>$.

Assuming about equal weak amplitudes for the two resonance states, one
estimates from (\ref{res})(\ref{g}) \cite{LASS, Anis}
\beqn
\frac{|A(1950, K^-\pi^+)|}{|A(1430, K^-\pi^+)|}&\approx &
\sqrt{\frac{[(m^2(D^0)-m^2)^2+m^2\Gamma^2] m'^2\Gamma' f'p_\pi}
{[(m^2(D^0)-m'^2)^2+m'^2\Gamma'^2] m^2\Gamma fp'_\pi}}= 2.1-2.4~,\nonumber\\
f' &  \equiv & {\rm BR}(K^*(1950)\to K^-\pi^+) = 0.35~,~~~
p'_\pi=904~{\rm MeV}~,
\eeqn
depending somewhat on $m'$ and $\Gamma'$. Namely, in the absence of a
radial suppression of its weak coupling to $D$, the resonance around 1900 MeV
contributes about 70$\%$ of the $I=1/2$ $D\to K\pi$
amplitude. In reality the contribution may be somewhat (but not very much)
smaller.

The combined contribution of the two resonances, at 1430 MeV and in the range
1820$-$1945 MeV, is considerably larger than the $I=3/2$ amplitude
in $D\to K\pi$. These contributions dominate the $I=1/2$ amplitude
if the two resonances interfere constructively.
This is the case if the mass of the second resonance is lower than
$m_D$, as claimed in \cite{Anis}. This explains the
$I =1/2$ dominance observed in these decays. In view of its
important role in $D$ decays, it would be helpful to determine
the mass of the higher resonance more precisely.

The above calculations show that direct channel resonances have
very large contributions in certain two body $D$ decays.
In a four-quark operator language (or in a diagram language) these
contributions are manifestations of annihilation (or W-exchange) amplitudes.
A possible phenomenological way of incorporating them in $D$ decays is by
employing a diagramatic language \cite{diag}, decomposing the $D\to K\pi$
amplitudes, for instance, into a color favored ``tree'' amplitude $T$, a
``color-suppressed'' amplitude C and an ``exchange'' amplitude $E$. In
a more general context this description is based on flavor SU(3) \cite{GHLR}.
Here we only assume isospin symmetry. The three amplitudes $T,~C,~E$ are
an over-complete set. Only two combinations are required to describe the two
isospin amplitudes
\beqn\label{isospin}
3A_{1/2} &=& 2T - C + 3E~,\nonumber\\
3A_{3/2} &=& T + C~.
\eeqn
The amplitude $T$ may be chosen to be real, $C$ obtains a complex phase from
rescattering, while $E$ is given by the sum of two Breit-Wigner forms,
representing the two resonances in $D\to K\pi$.

Clearly this scheme, which is more appropriate for the case of
large resonance contributions, deviates substantially from
the ``generalized factorization'' framework \cite{BSW, NS, C}. In the
latter prescription one combines the real amplitudes
\beqn\label{fact}
T &=& \frac{G_F}{\s}|V_{ud}V_{cs}|a_1f_\pi(m^2_D - m^2_K)F^{DK}(m^2_\pi)~,
\nonumber\\
C &=& \frac{G_F}{\s}|V_{ud}V_{cs}|a_2f_K(m^2_D - m^2_\pi)F^{D\pi}
(m^2_K)~,\nonumber\\
E &=& 0~,
\eeqn
into isospin amplitudes (\ref{isospin}) which are assigned arbitrary phases.
A large nonzero $E$ term, which is required in order to describe resonating
amplitudes, modifies the values obtained from the experimental data for
$a_1$ and $a_2$ relative to their values in the generalized factorization
prescription. Although the numerical changes may not be very large, which
is the reason for the {\it apparent success} of the generalized factorization
approach, the difference between the physical interpretations of the two
descriptions, with and without the $E$ term, is evident.

A fit of $D$ decays to $\bar K\pi,~\bar K\eta,~\bar K\eta'$ in terms of
diagrammatic amplitudes, assuming flavor SU(3) by which $T,~C,~E$ can be
separated, was carried out recently by Rosner \cite{R2}. He finds (in units
of $10^{-6}$~GeV)~$|T|\simeq 2.7,~|C|\simeq 2.0,~ |E|\simeq 1.6$.
A large phase ($-114^{\circ}$)
is found in $E/T$. The large magnitude of $E$,
comparable to the other two amplitudes, and its sizable phase
relative to $T$, are evidence for the important role of resonances in
these decays.

To demonstrate the insensitivity of the naive factorization prescription to
large nonfactorizable resonant contributions, we note the folowing:
Extracting $a_1$ and $a_2$ from the above values of $|T|$ and
$|C|$, using in (\ref{fact}) the values $F^{DK}(m^2_\pi)=0.77$~\cite{CLEO2},
$F^{D\pi}(m^2_\pi)=0.70,~f_K=160$~MeV, gives $|a_1|=1.06 ,~|a_2|=0.64$.
These values do not differ by too much from  $a_1=c_1(m_c)=1.26,~
a_2=c_2(m_c)=-0.51$, obtained in the traditional way which disregards
resonance contributions \cite{BSW, NS, C}.

While intermediate resonances were shown here to be important in $D^0$ decays,
they do not contribute to Cabibbo-favored $D^+$ decays, where the final states
consisting of $\bar d s\bar d u$ are pure $I=3/2$. This can be a qualitative
explanation for the measured longer $D^+$ lifetime. A calculation of the
$D^+/D^0$ lifetime ratio, including resonance contributions in $D^0$ decay,
is a challenging task.

To conclude, we comment on the possible effect of direct channel resonances
on $D^0-\bar D^0$ mixing. Reasonably small SU(3) breaking in
$D$ decays to two pseudoscalar mesons was shown to enhance the mixing by
several orders of magnitudes relative to the short distance box diagram
contribution \cite{DDbar}.
The actual enhancement was argued to be much smaller when summing over all
decay modes, if a large energy gap existed between the charmed quark mass
and $\Lambda_{\rm QCD}$ \cite{Geor}.
Resonance statess close to the $D$ mass violate this assumption.
Moreover, resonant contributions lead to particularly
large SU(3) breaking between SU(3)-related $D$ decay
rates. For instance, mass and widths differences between $K\pi$ and $\pi\pi$
resonances show up as
large rate differences (when CKM factors are included), since direct channel
resonance amplitudes peak strongly when the resonance mass approaches the $D$
mass. This raises the possibility that SU(3) breaking in resonance amplitudes
enhances $D^0-\bar D^0$ mixing beyond predictions based on the contributions of
a few two body decays \cite{DDbar}. Such effects were discussed recently
in \cite{GoPet}, where it was noted that in the lack of information about
weak Hamiltonian matrix elements between a $D$ meson and the resonances,
some crude assumptions must be made. The authors assume vacuum saturation
for these matrix elements, implying that P-wave $0^+$ resonances (for which
the wave functions vanish at the origin) do not
contribute to $D^0-\bar D^0$ mixing. Our model-independent calculation finds a
large matrix element for the $0^+$ $K\pi$ resonance at $1430$, which
indicates that the mixing can indeed be larger
than estimated in \cite{DDbar}. This interesting possibility deserves further
studies.

\bigskip
{\it Acknowledgements}: I thank Yuval Grossman, Alex Kagan, David Leith,
Alexey Petrov, Dan Pirjol, Jon Rosner, Michael Scadron and Michael Sokoloff for
useful discussions. I am grateful to the SLAC
Theory Group for its very kind hospitality. This work was supported in part
by the United States $-$ Israel Binational Science Foundation under research
grant agreement 94-00253/3, and by the Department of Energy under contract
number DE-AC03-76SF00515.

\newpage
\def \ajp#1#2#3{Am.~J.~Phys.~{\bf#1}, #2 (#3)}
\def \apny#1#2#3{Ann.~Phys.~(N.Y.) {\bf#1}, #2 (#3)}
\def \app#1#2#3{Acta Phys.~Polonica {\bf#1}, #2 (#3)}
\def \arnps#1#2#3{Ann.~Rev.~Nucl.~Part.~Sci.~#1, #2 (#3)}
\def \cmp#1#2#3{Commun.~Math.~Phys.~{\bf#1}, #2 (#3)}
\def \ib{{\it ibid.}~}
\def \ibj#1#2#3{~{\bf#1}, #2 (#3)}
\def \ijmpa#1#2#3{Int.~J.~Mod.~Phys.~A {\bf#1}, #2 (#3)}
\def \ite{{\it et al.}}
\def \jmp#1#2#3{J.~Math.~Phys.~{\bf#1}, #2 (#3)}
\def \jpg#1#2#3{J.~Phys.~G {\bf#1}, #2 (#3)}
\def \mpla#1#2#3{Mod.~Phys.~Lett.~A {\bf#1}, #2 (#3)}
\def \nc#1#2#3{Nuovo Cim.~{\bf#1}, #2 (#3)}
\def \npb#1#2#3{Nucl.~Phys. B~{\bf #1}, #2 (#3)}
\def \pisma#1#2#3#4{Pis'ma Zh.~Eksp.~Teor.~Fiz.~{\bf#1}, #2 (#3) [JETP
Lett. {\bf#1}, #4 (#3)]}
\def \pl#1#2#3{Phys.~Lett.~{\bf#1}, #2 (#3)}
\def \plb#1#2#3{Phys.~Lett.~B {\bf #1}, #2 (#3)}
\def \pr#1#2#3{Phys.~Rev.~{\bf#1}, #2 (#3)}
\def \pra#1#2#3{Phys.~Rev.~A {\bf#1}, #2 (#3)}
\def \prd#1#2#3{Phys.~Rev.~D {\bf #1}, #2 (#3)}
\def \prl#1#2#3{Phys.~Rev.~Lett.~{\bf #1}, #2 (#3)}
\def \prp#1#2#3{Phys.~Rep.~{\bf#1}, #2 (#3)}
\def \ptp#1#2#3{Prog.~Theor.~Phys.~{\bf#1}, #2 (#3)}
\def \rmp#1#2#3{Rev.~Mod.~Phys.~{\bf#1}, #2 (#3)}
\def \rp#1{~~~~~\ldots\ldots{\rm rp~}{#1}~~~~~}
\def \stone{{\it B Decays}, edited by S. Stone (World Scientific,
Singapore, 1994)}
\def \yaf#1#2#3#4{Yad.~Fiz.~{\bf#1}, #2 (#3) [Sov.~J.~Nucl.~Phys.~{\bf #1},
#4 (#3)]}
\def \zhetf#1#2#3#4#5#6{Zh.~Eksp.~Teor.~Fiz.~{\bf #1}, #2 (#3) [Sov.~Phys.
- JETP {\bf #4}, #5 (#6)]}
\def \zpc#1#2#3{Zeit.~Phys.~C {\bf #1}, #2 (#3)}

\end{document}